\documentclass[11pt]{article}
\usepackage{times}
\usepackage{sidecap}
\usepackage{geometry}
\usepackage{floatrow}
\usepackage{float}
\usepackage{hyperref}
\usepackage{graphics,graphicx}
\usepackage{subcaption}
\usepackage{epsf,wrapfig}
\usepackage{booktabs}

\usepackage{natbib,bm}

\usepackage{paralist}
\let\olditem\item
\renewenvironment{thebibliography}[1]{%
  \section*{\refname}
  \let\par\relax
  \renewcommand{\item}[1][]{\olditem[\textbullet]}%
  \inparaenum}{\endinparaenum}

\newcounter{aff}

\DeclareGraphicsExtensions{.jpg,.pdf,.png,.eps}

\usepackage{caption}
\usepackage{xcolor}

\usepackage{tcolorbox}

\bibpunct{(}{)}{;}{a}{}{,}

\usepackage{soul}

\definecolor{darkcyan}{rgb}{0.0, 0.0, 0.7}

\definecolor{boxcolor}{RGB}{193, 229, 255}
\definecolor{figboxcolor}{RGB}{193, 229, 255}
\definecolor{boxcolorgray}{RGB}{227, 227, 227}
\definecolor{darkcyan}{rgb}{0.0, 0.0, 0.7}

\sethlcolor{boxcolorgray}

\def\gtrsim{\lower 2pt \hbox{$\, \buildrel {\scriptstyle >}\over
{\scriptstyle \sim}\,$}}
\def\lesssim{\lower 2pt \hbox{$\, \buildrel {\scriptstyle <}\over
{\scriptstyle \sim}\,$}}

\def\approxlt{\lower.2em\hbox{$\buildrel < \over \sim$}}
\def\approxgt{\lower.2em\hbox{$\buildrel > \over \sim$}}

\def\be{\begin{equation}}
\def\ee{\end{equation}}
\def\bea{\begin{eqnarray*}}
\def\eea{\end{eqnarray*}}

\def\jref@jnl#1{{\rm#1}}
\def\aj{\jref@jnl{AJ}}			
\def\araa{\jref@jnl{ARA\&A}}		
\def\apj{\jref@jnl{ApJ}}			
\def\apjl{\jref@jnl{ApJ}}		
\def\apjs{\jref@jnl{ApJS}}		
\def\ao{\jref@jnl{Appl.~Opt.}}		
\def\apss{\jref@jnl{Ap\&SS}}		
\def\aap{\jref@jnl{A\&A}}		
\def\aapr{\jref@jnl{A\&A~Rev.}}		
\def\aaps{\jref@jnl{A\&AS}}		
\def\azh{\jref@jnl{AZh}}			
\def\baas{\jref@jnl{BAAS}}		
\def\jrasc{\jref@jnl{JRASC}}		
\def\memras{\jref@jnl{MmRAS}}		
\def\mnras{\jref@jnl{MNRAS}}		
\def\pra{\jref@jnl{Phys.~Rev.~A}}	
\def\prb{\jref@jnl{Phys.~Rev.~B}}	
\def\prc{\jref@jnl{Phys.~Rev.~C}}	
\def\prd{\jref@jnl{Phys.~Rev.~D}}	
\def\pre{\jref@jnl{Phys.~Rev.~E}}	
\def\prl{\jref@jnl{Phys.~Rev.~Lett.}}	
\def\pasp{\jref@jnl{PASP}}		
\def\pasj{\jref@jnl{PASJ}}		
\def\qjras{\jref@jnl{QJRAS}}		
\def\skytel{\jref@jnl{S\&T}}		
\def\solphys{\jref@jnl{Sol.~Phys.}}	
\def\sovast{\jref@jnl{Soviet~Ast.}}	
\def\ssr{\jref@jnl{Space~Sci.~Rev.}}	
\def\zap{\jref@jnl{ZAp}}			
\def\nat{\jref@jnl{Nature}}		
\def\iaucirc{\jref@jnl{IAU~Circ.}}
\def\aplett{\jref@jnl{Astrophys.~Lett.}}
\def\apspr{\jref@jnl{Astrophys.~Space~Phys.~Res.}}
\def\bain{\jref@jnl{Bull.~Astron.~Inst.~Netherlands}}
\def\fcp{\jref@jnl{Fund.~Cosmic~Phys.}}
\def\gca{\jref@jnl{Geochim.~Cosmochim.~Acta}}
\def\grl{\jref@jnl{Geophys.~Res.~Lett.}}
\def\jcp{\jref@jnl{J.~Chem.~Phys.}}	
\def\jgr{\jref@jnl{J.~Geophys.~Res.}}	
\def\jqsrt{\jref@jnl{J.~Quant.~Spec.~Radiat.~Transf.}}
\def\memsai{\jref@jnl{Mem.~Soc.~Astron.~Italiana}}
\def\nphysa{\jref@jnl{Nucl.~Phys.~A}}
\def\physrep{\jref@jnl{Phys.~Rep.}}
\def\physscr{\jref@jnl{Phys.~Scr}}
\def\planss{\jref@jnl{Planet.~Space~Sci.}}
\def\procspie{\jref@jnl{Proc.~SPIE}}

\def\ion#1#2{#1$\;${\small\rm\@Roman{#2}}\relax}

\newcount\lecurrentfam
\def\LaTeX{\lecurrentfam=\the\fam \leavevmode L\raise.42ex
\hbox{$\fam\lecurrentfam\scriptstyle\kern-.3em A$}\kern-.15em\TeX}
\def\sizrpt{
(\fontname\the\font): em=\the\fontdimen6\font, ex=\the\fontdimen5\font
\typeout{
(\fontname\the\font): em=\the\fontdimen6\font, ex=\the\fontdimen5\font
}}
\definecolor{dark}{rgb}{ 0.34, 0.09, 0.49}

\usepackage{fancyhdr,url}
\usepackage{transparent}
\usepackage{tablefootnote}
\usepackage{mathpazo}

\begin{document}

\noindent
\begin{raggedright}
{\Huge\bfseries Millimetre/Sub-Millimetre SETI\\[0.5em]
\Large Expanding the Radio Technosignature Parameter Space 
with ALMA2040
}\end{raggedright}\\

\noindent

Louisa A. Mason$^1$, Michael A. Garrett$^{1,2,3}$, Sofia Z. Sheikh$^4$, Kelvin Wandia$^1$, Anita M. S. Richards$^1$, Vishal Gajjar$^{4,6}$, Chenoa D. Tremblay$^{4,5,6,7,8}$, Andrew P. V. Siemion$^{1,3,4,5,6}$

\vspace{1em}
{
\footnotesize{
\noindent $^{1}$Jodrell Bank Centre for Astrophysics, Department of Physics and Astronomy, Alan Turing Building, University of Manchester, Oxford Road, M13 9PL, UK\\
$^{2}$Leiden Observatory, Leiden University, P.O. Box 9513, NL-2300 RA Leiden, The Netherlands\\
$^{3}$University of Malta, Institute of Space Sciences and Astronomy, Msida, MSD2080, Malta\\
$^{4}$SETI Institute, 339 Bernardo Avenue, Suite 200, Mountain View, CA 94043, USA\\
$^{5}$Astrophysics Sub-Department, Department of Physics, University of Oxford, Denys Wilkinson Building, Keble Road, Oxford OX1 3RH, UK\\
$^{6}$Berkeley SETI Research Center, University of California, Berkeley, CA 94720, USA\\
$^{7}$Department of Physics and Astronomy, University of New Mexico, Albuquerque, NM 87131, USA\\
$^{8}$National Radio Astronomy Observatory, 1003 Lopezville Rd., Socorro, NM 87801, USA\\
}
}

\begin{tcolorbox}[boxsep=3pt,left=0pt,right=0pt,top=1pt,bottom=1pt,colframe=boxcolor,colback=boxcolor]
\justify
\textbf{Abstract:}
{The search for extraterrestrial intelligence (SETI) has been largely concentrated at
centimetre wavelengths, and in particular in the `water hole' (1.42--1.66~GHz) range. The millimetre and
sub-millimetre sky is, by contrast, almost entirely unsearched. It
also offers some natural advantages: above 35~GHz, the interstellar and interplanetary media are effectively
transparent and non-dispersive, terrestrial interference is sparse, and interferometry suppresses
what interference remains. Both antenna gain and usable fractional
bandwidth increase with frequency, so a civilisation optimising directed, high-data-rate links has
advantageous reasons to operate in this regime, as illustrated by the ongoing migration of our own telecommunications systems to much higher frequencies. Building on the first ALMA technosignature survey
\citep{mason_conducting_2025}, we identify four searches ALMA2040 is uniquely placed to make:
(i) narrowband searches using a phased-up, beamformed array, extending the commensal model
pioneered for the VLA and MeerKAT into a regime where drift rates reach
$\sim10^{3}$~Hz~s$^{-1}$; (ii) broadband aggregate leakage (BRaTs) from Kardashev Type~I
civilisations, for which 128--512~GHz of instantaneous bandwidth places a $\sim60$~pc horizon
within an integration time of one hour on source; (iii) Dysonian searches for the cold waste heat of computation --- Slysh
haloes --- whose arcsecond and subarcsecond angular scales around M dwarfs are well matched to ALMA2040's
array configuration; and (iv) geometric discontinuities imprinted by occulting megastructures,
detectable as ringing in the visibility function with a phased ALMA operating in a mm/sub-mm VLBI
array. We set out the corresponding requirements: bandwidth,
beamforming, commensal access, spectral resolution and
sub-second time resolution.}
\end{tcolorbox}

\newpage{}   

\vspace{-0.7cm}
\section{Scientific context and open questions}

The search for extraterrestrial intelligence (SETI) is the differentiation of natural astrophysical emission and Earth-based interference from evidence of technological civilisations, known as technosignatures. A technosignature
must be unambiguous and distinct, and must indicate the presence of intelligent life or technology, past or present. The current resurgence in SETI {over the past decade has been driven by the availability of high-throughput digital backends that greatly increase the data volumes that can be searched} \citep{macmahon_breakthrough_2018, price_breakthrough_2018, tremblay_cosmic_2024, li_dual-backend_2026, tremblay_searching_2026}, wide-sky surveys conducted commensally \citep{tremblay_cosmics_2025, czech_breakthrough_2026}, and the employment of machine learning techniques for anomaly detection \citep{ma_deep-learning_2023, painter_novel_2025}.

The central challenge of SETI is a vast parameter space of unknown variables: we have no prior knowledge of how communication would occur, in what format, or with what purpose (intentional beacons or unintended leakage, for instance). SETI is therefore the pursuit of placing limits on a multidimensional parameter space spanning sky volume, frequency and bandwidth, sensitivity, temporal duration and periodicity, polarization and modulation \citep{wright_how_2018}. A null detection is pivotal to {constraining} this parameter space and evaluating the completeness of the search, and in turn the prevalence of extraterrestrial intelligent life within the cosmos.

Radio SETI remains at the forefront of technosignature research. Radio waves are a promising medium
for interstellar communication as they travel relatively unimpeded across galactic and
extragalactic scales, and are not absorbed by dust \citep{cocconi_searching_1959}. Electromagnetic
waves at radio frequencies remain favourable for Earth's own telecommunications and would
facilitate the fastest forms of communication for intentional broadcasts
\citep{tremblay_searching_2026}. In addition, coherent radio transmissions are energetically (and
hence economically) efficient to produce, and large bandwidths
(especially at mm and sub-mm wavelengths) can be used to convey high
informational data rates through free space \citep{manunza_first_2025, myburgh_vla_2026}.

A fundamental variable in the `cosmic haystack' is the choice of frequency. Early SETI efforts were predominantly
confined to L-band, concentrating on the window of low Galactic emission bounded by the neutral
hydrogen and hydroxyl lines between 1.42 and 1.67~GHz --- the `water hole'
\citep{OLIVER197971} --- as well as the sensitivity of receivers at the time. Recent efforts have expanded toward lower frequencies with the Murchison
Widefield Array e.g. \citep{tingay_opportunistic_2016} and LOFAR \citep{johnson_simultaneous_2023}, while progress toward higher frequencies has been more gradual. High-frequency SETI has been held back partly
because the specialised instrumentation and telescopes are rare and expensive, and building sensitive high-frequency receivers can be challenging \citep{myburgh_vla_2026}. To date, only a few demonstrations have been carried out: a search targeting the 203~GHz positronium hyperfine resonance \citep{steffes_seti_1994}, a 30-target survey with the Sardinia Radio Telescope at 18~GHz \citep{manunza_first_2025}, the COSMIC 25--50~GHz search with the VLA \citep{myburgh_vla_2026}, and the first ALMA technosignature survey, conducted at 90~GHz on archival data \citep{mason_conducting_2025}. Collectively, these searches cover a vanishingly small fraction of the 35--500~GHz frequency range that ALMA2040 will access.

It is worth being explicit about \emph{why} a civilisation might transmit in this region, since the
familiar arguments for the water hole do not apply. First, for a fixed aperture, the gain of a transmitting antenna scales as
$\nu^{2}$, so the effective isotropic radiated power (EIRP) achievable from a given transmitter
power and dish rises quadratically with frequency: mm-wave
transmitters are good for directed, high-EIRP links. Second, for a fixed fractional bandwidth, usable bandwidth scales with carrier frequency:
for an approximation, 3~GHz is 1\% of 300~GHz, sufficient for data rates far beyond anything available at
L-band, but greater fractional bandwidths may be more realistic but increasingly intensive to search. Third, the mm/sub-mm band is essentially immune to the propagation effects that limit
lower-frequency SETI. In particular, temporal and spectral broadening by the ionised interstellar and
interplanetary media scale steeply with wavelength and are negligible above 35~GHz, except in the densest plasma density regions \citep{mason_conducting_2025} , allowing intrinsically narrowband signals with milli-Hz bandwidths to remain narrow upon arrival. Interstellar scintillation,which imposes stochastic amplitude variability on compact sources at centimetre wavelengths \citep{garrett_isolating_2026}, is also correspondingly suppressed. The fourth argument is empirical: our own telecommunications
infrastructure is migrating upwards, from the 5G mm wave allocations at 24--47~GHz towards sub-THz carriers proposed for 6G. Extrapolating our present-day L-band leakage spectrum to a Type~I civilisation may therefore substantially understate what such a civilisation radiates at
tens to hundreds of GHz.

The honest counterweight is that mm/sub-mm is not the radiometrically quiet part of the spectrum. The sky background reaches its minimum at a few GHz; above it, the cosmic microwave
background's specific intensity, atmospheric emission, and the receiver quantum limit 
all rise \citep{mason_conducting_2025}. Furthermore, ALMA's predominant constraint in noise remains the telluric sky lines. However, what the mm/sub-mm band offers is not a low system temperature but a low density of \emph{confusing structure}, and this is the claim that matters for SETI. 

Narrowband astrophysical confusion is minimal. Masers produce narrow lines, typically corresponding to kHz, and some lines contain hyperfine structure (although this is rarely resolved spectrally). 
    The narrowest natural feature known anywhere in the radio spectrum --- structure in OH maser components --- is $\sim500$~Hz wide \citep{cohen_narrow_1987}. Although frequencies of some mm-wave transitions are only catalogued to a few MHz precision, maser lines bright enough to be detectable in sub-kHz channels are likely to be well known and occur alongside other species characteristics of the environment providing the pumping conditions. 
    A significant fraction of the population of rest frequencies are known to high precision from laboratory spectroscopy, so the candidate false positives can largely be looked up rather than characterised empirically, which is not true of terrestrial RFI at L-band. 
    The mm/sub-mm regime is also comparatively free
    of the non-thermal processes that dominate at lower frequencies, with synchrotron emission
    falling steeply towards these wavelengths \citep{kardashev_optimal_1979}.
    
The confusing continuum population changes character with respect to frequency, rather than disappearing. The
    synchrotron dominated sources that limit deep cm-wave surveys fade towards mm wavelengths,
    but the faint sub-mm sky is instead populated by high-redshift dusty star-forming galaxies
    - integrating deeper does not thin them out. Deep continuum surveys resolve a
    growing fraction of the cosmic infrared background into individual $\mu$Jy sources, and these
    are already a recognised contaminant of Dysonian candidate lists
    \citep{ren_high-resolution_2025}. Any sub-mm technosignature deep field is therefore
    confusion-limited by galaxies, not by an empty sky. 
    Because these galaxies sit at cosmological distances and are resolved on scales of a few tenths of an arcsecond, an extended configuration can resolve them out, while astrometry separates them from a source co-moving with a Galactic star \citep{garrett_isolating_2026}.

 The RFI environment is quieter at mm/sub-mm. Interferometry further suppresses RFI at any frequency, since
    terrestrial interference does not correlate on long baselines; spatial information provided by interferometers, with or without correlation, makes a significant difference in RFI discrimination. In addition, an increased number of baselines strengthens confidence in a detection, offering redundancy
    in the case of a non-repeating transmission \citep{garrett_seti_2018,wandia_interferometric_2023}.

\begin{figure}
    \centering
    \includegraphics[width=\linewidth]{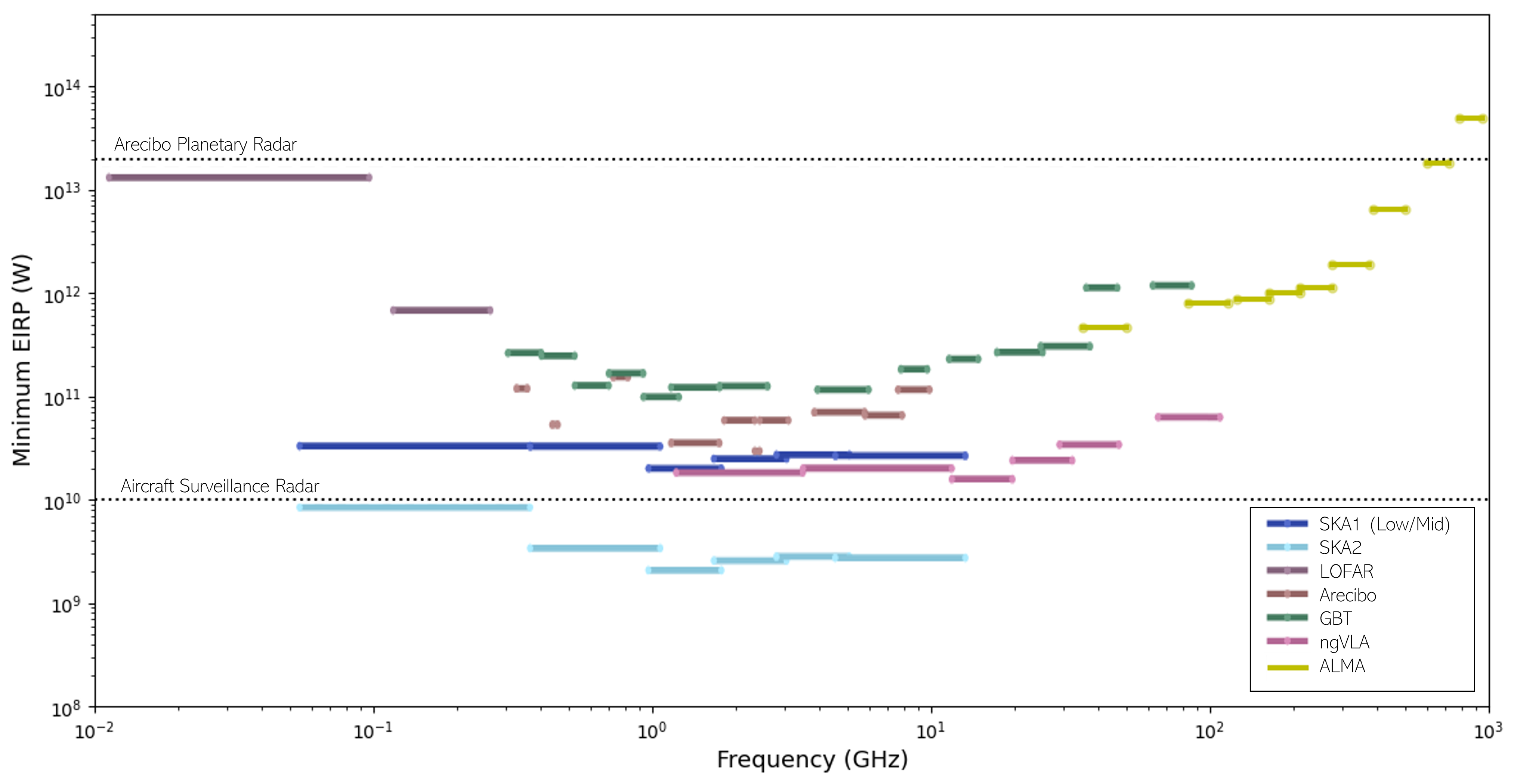}
    \caption{{Comparison of the frequency coverage and minimum detectable EIRP of facilities
    used for technosignature surveys \citep{croft_science_2018, mason_conducting_2025}. Estimates of EIRP are based from an integration time of 10 min, a significance threshold of 15, channel width of 0.5 Hz, and distances up to 15 pc. ALMA occupies a
    frequency range that no other current SETI facility probes.}}
    \label{fig:EIRP_facilities}
\end{figure}

Because interferometry suppresses the few remaining sources of confusion across the mm/sub-mm spectrum, the primary limitation for these SETI surveys becomes the sensitivity of the instrument.
Figure \ref{fig:EIRP_facilities} compares a sample of instruments currently used for technosignature
surveys, plotting frequency coverage against minimum EIRP sensitivity. Notably, ALMA explores a high-frequency parameter space
that remains otherwise unexploited for SETI. However, on raw narrowband EIRP, it is also the least sensitive facility on
the plot: current ALMA configurations reach $\rm{EIRP_{min}}\sim10^{12}$--$10^{13}$~W above 100~GHz, compared to $\sim10^{10}$~W for SKA at L-band \citep{croft_science_2018}. Highlighting this gap defines what the ALMA2040 upgrade must deliver. The case for mm/sub-mm SETI
rests on (i) covering parameter space that no other facility can reach, and (ii) the fact that three of the four signatures discussed below---broadband leakage, cold Dysonian waste heat, and occultation ringing---are \emph{intrinsically}
mm/sub-mm phenomena that do not compete with narrowband cm-wave searches at all. ALMA2040 will
close this gap through bandwidth expansions. A 17--68x increase in instantaneous bandwidth
over current ALMA translates into a 4--8x improvement in continuum sensitivity, alongside a 5--10$\times$ improvement in line sensitivity that yields a comparable gain for narrowband observations. The four avenues are:

\textbf{Narrowband transmissions} have been the simplest form of technosignature to
    search for, given that an artificial intentional broadcast with a bandwidth on Hz scales is
    distinct from the narrowest natural emission {\citep{cohen_narrow_1987}}. However, there
    have been reservations about searching for narrowband transmissions at mm/sub-mm wavelengths.
    The drift rate of a narrowband transmission, due to the relative acceleration between
    a transmitter and receiver, scales multiplicatively with frequency, so a narrowband signal can rapidly `blur' across many frequency channels \citep{sheikh_choosing_2019, margot_search_2021}. Adopting the
    Breakthrough Listen convention of $\pm4$~Hz~s$^{-1}$ at 1~GHz {gives maximum drift rates of
    $\pm160$~Hz~s$^{-1}$ in Band~1 and $\pm920$~Hz~s$^{-1}$ in Band~6.} Since mm/sub-mm facilities
    have a typical spectral resolution of 30~kHz, {a drifting signal nonetheless remains within
    a single channel for 75~s in Band~3 and 33~s in Band~6}, so Doppler blurring may be
    compensated for by integrating on timescales of minutes \citep{mason_conducting_2025}.
    {Recovering the full sensitivity of a genuinely Hz-wide carrier requires finer
    channelisation and an explicit de-Doppler search; the implications for the ALMA2040 backend are
    quantified in Section~\ref{sec:beamforming}.}

    The reduced field of view of mm/sub-mm facilities has historically restricted large scale radio SETI surveys  that aim to maximize target volume. For ALMA's current
    bandwidth and correlated integration time, bandwidth and time-average smearing do not reduce
    the field of view below the primary beamwidth at any frequency \citep{mason_conducting_2025}.
    Furthermore, strategies to maximise the `stellar bycatch' by estimating incident peripheral
    stellar populations can greatly increase the limits placed by a single survey. Using
    astrometric catalogues such as \textit{Gaia} increases the population of stellar objects
    considered by $10^2$ \citep{wlodarczyk-sroka_extending_2020}, and recent work using Galactic
    simulation models increases it by an additional order of magnitude \citep{mason_simulating_2026}.
    Considering the stellar bycatch not only maximises an otherwise limited field of view, but
    expands the breadth of spectral types considered that would otherwise be removed through
    classification and binary confusion \citep{mason_simulating_2026}. {This is the principal
    lever by which a small field of view is made to yield a large
    survey, and it may argue for commensal rather than targeted
    operation in most search scenarios (Section~\ref{sec:commensal}).}

\textbf{Broadband leakage} is the signature of a civilisation that is not trying to
    be found.
    {Terrestrial technology has evolved from a small number of
    powerful narrowband transmitters towards a vast, heterogeneous population of low-power
    broadband emitters: cellular base stations, mobile devices, WiFi, satellite constellations,
    radar and navigation systems. The aggregate of millions of such incoherent emitters distributed
    across a planetary surface and in low orbit produces a spectrally extended, faint continuum
    signal, easily mistaken for unstructured noise by conventional SETI analyses}
    \citep{garrett_isolating_2026}. {These broadband radio technosignatures (BRaTs) from an intermediate  Kardashev Type I/Type II civilisation --- one consuming up to the equivalent power of its host star,
    $\sim10^{16}$~W --- are estimated to have a leakage spectral luminosity
    $L_{\nu}\approx4\times10^{5}$~W~Hz$^{-1}$ at L-band, corresponding to $\sim33~\mu$Jy at 10~pc and
    $\sim1~\mu$Jy at 60~pc \citep{garrett_isolating_2026}.}

    {BRaTs have not previously been considered at mm wavelengths, but the case is
    promising. They are, by construction, insensitive to Doppler drift, which removes the need for
    fine channelisation and permits arbitrarily long coherent integration, providing the opportunity for a ``SETI Deep
    Field'' approach \citep{garrett_isolating_2026}. That is exactly the regime in which ALMA2040's
    bandwidth is decisive. The projected ALMA2040 Band 1 point-source sensitivity is
    $0.81~\mu$Jy~beam$^{-1}$ per hour over 7.5~GHz. Scaling this to 128~GHz of
    instantaneous bandwidth improves the hourly sensitivity to $0.20~\mu$Jy~beam$^{-1}$, while expanding to 512~GHz via simultaneous multi-band observing yields $0.10~\mu$Jy~beam$^{-1}$. The corresponding Type~I leakage horizons are $\sim60$~pc for a single hour integration and $\sim260$~pc within a 100-hour deep field, compared to the $\lesssim100$~pc quoted for SKA-era cm-wave arrays \citep{garrett_slysh_2026} This horizon assumes the caveat that the aggregate leakage spectral luminosity at 40~GHz is comparable to its L-band value.
    That assumption is untested, but the upward migration of our own spectrum within several decades motivates the use of ALMA2040 as the only instrument that could test it.}

\textbf{Dysonian SETI} expands beyond using infrared to identify excess heat candidates.
    Radio may be fundamental to the exploration of these technosignatures, not only to
    cross-examine candidates for contamination by dust-obscured galaxies
    \citep{ren_high-resolution_2025}, but to explore novel signatures of {mature} technological
    civilisations as well. One such signature originates from cold computation, radiating as grey thermal
    emission at the local equilibrium temperature \citep{garrett_slysh_2026}. {The Landauer
    cost of irreversible computation scales linearly with temperature \citep{laundauer_computing}, so a computation-dominated
    civilisation has a thermodynamic incentive to operate cold; because the required radiating area
    then scales as $T^{-4}$, cold computation is conspicuous rather than hidden.} These
    \textit{Slysh Haloes} are predicted to occupy the 5--28~K zone of a planetary system, with M
    dwarfs providing the most favourable hosts observationally. The two components (star and halo) are
    mutually invisible: the stellar photosphere is faint in the sub-millimetre, while the halo
    emits almost nothing in the optical \citep{garrett_slysh_2026}.

    {The angular-scale argument makes the case for ALMA2040 and other interferometric facilities, over
    single-dish facilities. A 12-K halo around Proxima Centauri would span
    $\sim33''$ and would be resolved out
    by any interferometer; such nearby systems belong to JCMT and \textit{Planck}. Halo radius
    scales as $L_{*}^{1/2}$ and angular size as $d^{-1}$, so an M5V halo at 100~pc subtends only
    $\sim0.4''$ across its 12-K zone, with the full 5--28~K region spanning $0.04''$--$1.2''$. Both
    fall comfortably inside ALMA2040's most compact configuration, which delivers $0.45''$ resolution
    and a $3.6''$ maximum recoverable scale in Band~7. 
    
    ALMA2040 extends the horizon for compact haloes from the ~10 pc reachable at SCUBA-2-class survey depths (Garrett 2026a) to ~100 pc — a factor ~10 in distance, and hence ~10³ in searched volume. More importantly, ALMA2040 \emph{resolves} candidates, enabling
    their distinction from natural debris discs or cold cores. Furthermore, 
    multi-band imaging turns a detection into a measurement
    of the radial distribution of the emitting material.}

    Another strategy to explore Dysonian SETI at mm/sub-mm is to search for distinct morphological signature that could arise from occultation. An occulting extraterrestrial
    megastructure would produce an artificial sharp edge in front of a background radio source.
    These geometric discontinuities produce `ringing' in the visibility function, distinguishing
    them from typical astrophysical sources (Mason et al., in preparation). {The relevant scale is set by the array, not by
    the structure: a mm/sub-mm VLBI array with $10^{4}$~km baselines resolves $\sim20~\mu$as at
    345~GHz, corresponding to $\sim4\times10^{4}$~km at 10~pc --- comparable to the Fresnel scale on which a sharp edge imprints its diffraction pattern. A phased ALMA is the dominant sensitivity element of any
    such array, and the signature is therefore uniquely accessible to a phased-up ALMA2040}

 \textbf{Transient/Periodic pulsed beacons} offer an energetically inexpensive means of
    transmission compared with broadband or long-duty-cycle narrowband continuous-wave
    transmissions \citep{suresh_48_2023}. At mm/sub-mm wavelengths, we don't expect significant natural dispersion for most Galactic sources, and so signals with significant dispersion may be significant for technosignatures \citep{gajjar_breakthrough_2021, gajjar_searching_2022}. Periodic signals are currently searchable only on
    timescales longer than ALMA's 6~s correlator dump, which excludes the entire millisecond-to-second
    regime in which terrestrial pulsed systems and known astrophysical analogues operate.
    Sub-second dump times, or access to beamformed voltage products, would open it. Transient
    surveys conducted for SETI offer complementarity with pulsar detections
    \citep{perez_deepest_2026}. To detect periodic technosignatures, a fast folding algorithm
    (FFA) can be applied per channel, as in the BLIPSS pipeline \citep{suresh_48_2023}.
    Distinction from rapid transient fluctuations in astrophysical emissions at mm would be necessary, in the case of cyclotron fluctuations (on the order of hours) and rapidly-varying stellar phenomena amplified by masers (on the order of minutes) \citep{das_discoveries_2025, fujisawa_fast_2026}. Despite limited investigation into the  hyperfine structure and variability at mm/sub-mm wavelengths, we should be vigilant for potential SETI confusion without assuming the effects remain at cm; the technical methods proposed for SETI will also open up astrophysical studies, offering synergy between SETI and other astrophysics fields.  

The pursuit of technosignatures, despite presently being a largely unexplored field at mm/sub-mm wavelengths, falls naturally within the science case of ALMA2040. 
ALMA2040 will characterise the birth of planetary systems and their atmospheric chemistry with unprecedented fidelity; the same data stream, at no additional observing cost, can be interrogated for evidence that one of them became technological.

\section{ALMA2040: Technical Requirements}

Radio SETI searches use channelized spectra derived from single-dish or beamformed interferometric voltage data, applying de-Doppler algorithms to identify candidate signals, or ‘hits’. Systems such as COSMIC perform these searches in real time and record hit data for subsequent analysis.
Alternatively, interferometers may be employed in image-plane SETI. Since the position of a star or signal is invariant and localised within the field of view \citep{garrett_seti_2018, tremblay_seti_2020}, it is possible to slice through the data cube for a given pixel across frequency; extraterrestrial artificial emission would deviate from the Gaussian distribution of random noise across the cube. 

\subsection{A phased-up ALMA2040: beamformed narrowband SETI}
\label{sec:beamforming}

We consider ALMA2040 with approximately 250 antennas, 128~GHz of instantaneous
bandwidth, and the 1~Hz channelisation that narrowband SETI considers ideal. This yields $1.28\times10^{11}$
channels across $N(N-1)/2 = 31{,}125$ baselines, presenting a massive computational problem. Fortunately, the field has already converged on a solution at cm wavelengths \citep{tremblay_cosmic_2024, tremblay_searching_2026}. \emph{We propose
that ALMA2040 be designed from the outset as a phased array capable to run in interferometric mode, 
beamforming mode or, critically, both simultaneously.} ALMA already possesses a beamforming capability through the ALMA Phasing System \citep{matthews_alma_2018}, which was developed to incorporate ALMA as an element of
the Event Horizon Telescope; ALMA2040 should treat this not as a specialist VLBI add-on, but as a first-class observing mode supported by a SETI-capable backend.

{In beamforming mode, the two beam types serve different purposes. \emph{Coherent} beams sum the antenna voltages in phase, delivering the full collecting area of the array within a beam of the synthesised size. Their sensitivity scales as $N$, allowing for high-resolution view of multiple object simultaneously within the telescope's field of view. Conversely, \emph{incoherent} beams sum the antenna powers, covering the entire primary beam at a sensitivity penalty of $\sqrt{N}$ (a factor 15.8 for 250 antennas); they are the more suited for capturing stellar bycatch across the full field of view. Both are needed.}

{The number of coherent beams required to tile the primary beam is independent of frequency, since both the primary beamwidth and the synthesised beam scale as $\nu^{-1}$. Full-field coherent beamforming becomes increasingly challenging with high number of antennas and longer baselines, resulting in narrower beams. This points to a clear conclusion: \textbf{beamformed SETI is best suited to the most compact and intermediate configurations}, and we must prioritize targets as a result. While full-field coherent beamforming should be feasible with future GPU technology, tiling the entire primary beam becomes progressively less practical in the most extended configurations.}

{Drift rate is the second design driver, and its impact is severe. With $\dot{\nu}_{max}=\pm4\nu$~Hz~s$^{-1}$
($\nu$ in GHz), a 2~Hz channel typically used at 25--50~GHz \citep{myburgh_vla_2026} is
coherent for only 12~ms in Band~1 and 2~ms in Band~6. A de-Doppler search must therefore cover a drift range that grows linearly with frequency, which causes the number of acceleration trials to scale accordingly. This has a concrete architectural consequence: full-bandwidth, Hz-resolution SETI data cannot be stored, even in beamformed format. The search must therefore be performed \emph{in real time}, with only detected candidates and their associated voltage `stamps' written to disk. This is the exact model currently established at the VLA and MeerKAT
\citep{tremblay_cosmics_2025, czech_breakthrough_2026}, and it should be treated as a design input to ALMA2040 rather than something retrofitted afterwards. We further recommend that the backend support \emph{tunable} channelisation between $\sim1$~Hz and $\sim30$~kHz, allowing a survey to trade drift-rate coverage against spectral dilution on a band-by-band basis.}

{This setup provides a direct scientific continuation of existing searches. The COSMIC 25--50~GHz survey \citep{myburgh_vla_2026}---the highest-frequency large-scale beamformed narrowband search published to date---reached an $EIRP_{min}\sim10^{15}$--$10^{17}$~W towards 45 \textit{Gaia} targets, using
2~Hz channels and a $\pm50$~Hz~s$^{-1}$ drift search. Its upper frequency of 50~GHz lies inside
ALMA Band~1 (35--50~GHz). A beamformed ALMA2040 picks up precisely where the VLA's Q-band leaves off, with an order of magnitude more collecting area and two orders of magnitude more bandwidth, and carries the search continuously towards 500~GHz.}

\subsection{Commensal operation}
\label{sec:commensal}

{ALMA2040's field of view is small --- $157''$ in Band~1, $27''$ in Band~6. Commensal
operation is the answer, and it is proven.} Arrays constructed with Ethernet-based architecture and
multicasting capabilities enable the splitting, or `carbon copy', of an observation before it is
passed to a backend. This strategy has been effective with MeerKAT, where the BLUSE system has captured 1.2 million pointings since 2022
\citep{czech_breakthrough_2026}. There is a strong case for future phased-array instruments to
adopt a ``User Supplied Equipment'' approach, enabling broader science cases through the
installation of multiple backends in parallel, as demonstrated with MeerKAT
\citep{slabber_meerkat_2018, czech_breakthrough_2026}. For high-throughput facilities, raw voltages are stored only for
signals of interest, as `stamps', rather than storing the entire data stream
\citep{tremblay_searching_2026}. Careful consideration of RFI is required when building such
pipelines at L-band, but this is far less contentious at mm/sub-mm wavelengths.

{Combined with the stellar bycatch methodology \citep{wlodarczyk-sroka_extending_2020,
mason_simulating_2026}, commensal access converts ALMA2040's entire observing programme into a
technosignature survey at essentially zero cost in observing time --- the single most
cost-effective SETI capability that could be built into the facility. Concretely, this requires:
(i) an Ethernet-based digital backend with multicast output; (ii) a defined User Supplied
Equipment interface and the operational policy to permit it; and (iii) the ability to run a
beamformer commensally with interferometric observing, so that SETI beams are formed on
catalogued stars within whatever field the primary observer has chosen.}

\begin{table}[h]
\centering
\footnotesize
\caption{{ALMA2040 capabilities and the technosignature science they enable, in approximate
priority order for this science case.}}
\label{tab:reqs}
\begin{tabular}{@{}p{3.5cm}p{4.2cm}p{6.6cm}@{}}
\toprule
\textbf{Capability} & \textbf{Requirement} & \textbf{SETI driver} \\
\midrule
Instantaneous bandwidth & $\geq128$~GHz; goal 512~GHz via simultaneous multi-band &
Sets the BRaTs horizon ($\sim60$~pc~hr$^{-1}$); removes the anthropocentric `magic frequency'
guess by covering 35--1000~GHz at once \\
\\
Beamforming backend & Coherent 
and incoherent beams; tunable 1~Hz--30~kHz channelisation; 
stamp recording &
The only tractable route to narrowband SETI at ALMA2040 data rates; enables mm-VLBI
(occultation ringing) \\
\\
Commensal access & Ethernet/multicast backend, User Supplied Equipment policy &
Converts the full observing programme into a survey; only way to accumulate sky coverage with a
$\lesssim157''$ field of view \\
\\
Line sensitivity & 5--10$\times$ improvement, prioritised for Band~7 and above &
Lowers $EIRP_{min}$ towards parity with cm-wave facilities; Slysh halo detection at 100~pc \\
\\
Angular resolution & 3--5$\times$ improvement; phased operation within a global mm-VLBI array &
Resolves Slysh halo radial structure; $\sim20~\mu$as VLBI resolution for occultation ringing and
sub-mas astrometric confirmation of BRaT candidates \citep{garrett_isolating_2026,
wandia_interferometric_2023} \\
\\
Time resolution & Sub-second dumps; access to beamformed voltages &
Opens the ms--s pulsed-beacon regime, currently excluded by the 6~s correlator dump \\
\\
250 antennas & --- &
$N$-fold coherent beam sensitivity; baseline redundancy gives independent confirmation of a
non-repeating detection \\
\\
Short spacings / total power & ACA-equivalent or single-dish combination &
Recovers the $>10''$ Slysh haloes of the very nearest stars, which an interferometer alone
resolves out \\
\bottomrule
\end{tabular}
\end{table}

We must consider any potential changes to the limits outlined in \citet{mason_conducting_2025} that also translate to
ALMA2040. Bandwidth smearing does not limit the field of view below the primary beamwidth, even
with improved spectral resolution. However, time-average smearing will halve the undistorted field
of view for the proposed most extended configuration (28~km); at the limit where time-average
smearing is equivalent to the primary beamwidth, the maximum baseline is 14.75~km, so intermediate
and compact configurations are unaffected. This reinforces the configuration argument of
Section~\ref{sec:beamforming}: \textbf{the compact and intermediate configurations are the natural home for
wide-field SETI, and the extended configuration is reserved for targeted, high-resolution follow-up
of specific candidates.} A summary of our arguments for ALMA2040's deployment for SETI can be found in table \ref{tab:reqs}.



\section{Synergies with current and future facilities}

Millimetre/sub-millimetre SETI is complementary to, not competitive with, the cm-wave programmes. SKA and
ngVLA will conduct the deep, wide, sub-$\mu$Jy continuum surveys within which BRaT candidates are
first identified \citep{garrett_isolating_2026, tremblay_searching_2026}; ALMA2040 supplies the
spectral lever arm that tests the engineered spectral non-uniformity --- absorption `notches',
regulated-allocation cut-offs --- on which the BRaTs diagnostic framework depends, and which cannot
be measured within a single decade of frequency. For Dysonian searches the division is by angular
scale: all-sky \textit{Planck} data and single-dish JCMT/AtLAST observations cover the arcminute
haloes of the nearest stars, while ALMA2040 resolves the arcsecond and subarcsecond  haloes of the far more numerous
systems at tens to hundreds of parsecs \citep{garrett_slysh_2026}. A phased ALMA2040 within the
EHT and its successors provides the $\mu$as-scale imaging needed for occultation and for
astrometric confirmation \citep{matthews_alma_2018, wandia_interferometric_2023}. Finally, this
white paper depends on capabilities --- bandwidth, beamforming, commensal backends, short spacings
--- that are shared with the exoplanet, protoplanetary disc and Galactic Centre cases; we
recommend explicit coordination with those working groups, since the technosignature case adds
essentially no new hardware requirement beyond a backend and an access policy.

\section{Conclusion}

ALMA2040 will be the most sensitive instrument ever built in a frequency range that
technosignature searches have barely touched. The case rests on four propositions. 

First, mm/sub-mm wavelengths are the only part of the radio spectrum in which interstellar
propagation is effectively perfect and natural narrowband emission is 
broader than an engineered carrier; the dominant continuum confounders --- high-redshift dusty galaxies ---
are not sparse, but they are cleanly separable by spectral index, angular size and astrometry.
Second, the physical incentives for a transmitting civilisation 
point upwards in frequency, as is our own spectrum usage. The idea that Type~I leakage is an L-band phenomenon is an
assumption, and ALMA2040 is the instrument that can test it, with a $\sim130$~pc horizon in an hour
and $\sim600$~pc in a deep field. Third, two technosignature classes --- the cold waste heat of
computation, and the diffraction signature of an engineered occulting edge --- are intrinsically
sub-millimetre phenomena with no cm-wave counterpart, and ALMA2040's angular resolution is matched
to both. Fourth, the capabilities required are modest and largely shared: an instantaneous
bandwidth above $\geq128$~GHz, a beamforming backend with 
stamp recording, commensal Ethernet access under a User Supplied Equipment policy, and sub-second time
resolution. 

{The absence of a detection is itself a result. Every hour ALMA2040 spends observing
protoplanetary discs, exoplanetary atmospheres and evolved stars is an hour in which, at no
additional cost, the same data can constrain the prevalence of technological life in a volume and a
frequency range that no other facility will reach. 

\section*{Acknowledgements}
The authors wish to thank reviewers Karen I. Perez, Yuri Uno and Laura Sommovigo for their comments. This work was developed as part of the broader ALMA2040 community effort. We acknowledge the
Lorentz Center and NOVA for supporting the community workshops and coordination activities that
helped initiate and shape the ALMA2040 White Paper process.\\

\bibliographystyle{apj_mod}
\bibliography{seti}
\end{document}